\documentstyle[12pt,epsf]{article}
\newcommand{\abstracts}[1]{{
\centering{\begin{minipage}{12.2truecm}
\normalsize\baselineskip=15pt
\centerline{\footnotesize ABSTRACT}\vspace*{0.3cm}
\parindent=20pt #1
\end{minipage}}\par}}
\newcommand{\beqn}{\begin{eqnarray}}
\newcommand{\eeqn}{\end{eqnarray}}
\newcommand{\eq}[1]{(\ref{#1})}
\newcommand{\beq}{\begin{equation}}
\newcommand{\eeq}{\end{equation}}
\newcommand{\dd}{\mbox{d}}

\def\bbbone{{\mathchoice {\rm 1\mskip-4mu l} {\rm 1\mskip-4mu l}
{\rm 1\mskip-4.5mu l} {\rm 1\mskip-5mu l}}}
\def\bbbc{{\mathchoice {\setbox0=\hbox{$\displaystyle\rm C$}\hbox{\hbox
to0pt{\kern0.4\wd0\vrule height0.9\ht0\hss}\box0}}
{\setbox0=\hbox{$\textstyle\rm C$}\hbox{\hbox
to0pt{\kern0.4\wd0\vrule height0.9\ht0\hss}\box0}}
{\setbox0=\hbox{$\scriptstyle\rm C$}\hbox{\hbox
to0pt{\kern0.4\wd0\vrule height0.9\ht0\hss}\box0}}
{\setbox0=\hbox{$\scriptscriptstyle\rm C$}\hbox{\hbox
to0pt{\kern0.4\wd0\vrule height0.9\ht0\hss}\box0}}}}
\def\bbbe{{\mathchoice {\setbox0=\hbox{\smalletextfont e}\hbox{\raise
0.1\ht0\hbox to0pt{\kern0.4\wd0\vrule width0.3pt
height0.7\ht0\hss}\box0}}
{\setbox0=\hbox{\smalletextfont e}\hbox{\raise
0.1\ht0\hbox to0pt{\kern0.4\wd0\vrule width0.3pt
height0.7\ht0\hss}\box0}}
{\setbox0=\hbox{\smallescriptfont e}\hbox{\raise
0.1\ht0\hbox to0pt{\kern0.5\wd0\vrule width0.2pt
height0.7\ht0\hss}\box0}}
{\setbox0=\hbox{\smallescriptscriptfont e}\hbox{\raise
0.1\ht0\hbox to0pt{\kern0.4\wd0\vrule width0.2pt
height0.7\ht0\hss}\box0}}}}

\begin{document}

\begin{flushright}
{\large ITEP-TH-48/97}
\end{flushright}
\begin{flushright}
{\large HUB-EP-97/68}
\end{flushright}
\vspace{.7cm}

\begin{center}

{\baselineskip=16pt
{\Large \bf Topological Content of the Electroweak Sphaleron}\\
{\Large \bf on the Lattice}

\vspace{1cm}

{\large M.~N.~Chernodub\footnote{e-mail:
chernodub@vxitep.itep.ru}$^{,a}$,
F.~V.~Gubarev\footnote{e-mail:
Fedor.Gubarev@itep.ru}$^{,a}$
and E.--M.~Ilgenfritz\footnote{e-mail:
ilgenfri@hep.s.kanazawa-u.ac.jp}$^{,b,c}$}\\

\vspace{.5cm}
{ \it
$^a$ ITEP, B.Cheremushkinskaya 25, 117259 Moscow, Russia \\

%\vspace{0.3cm}
$^b$ Institut f\"ur Physik, Humboldt--Universit\"at,
D--10115 Berlin, Germany \\
$^c$ Institute for Theoretical Physics, University of Kanazawa, \\
JP-920-1192 Kanazawa, Japan
}
}
\end{center}
\vspace{1cm}

\abstracts{
$3D$ electroweak sphalerons on the lattice
are used as test configurations for
definitions of various topological defects.
In the maximally Abelian gauge they are
shown to contain a symmetric array of Abelian monopoles and
anti-monopoles connected by two kinds of Abelian vortex strings.
Gauge--invariant lattice definitions of the Nambu monopole and
the $Z$--vortex string are formulated which correspond
to Abelian projection from the unitary gauge.
The sphalerons contain in their core just one (non--Abelian) Nambu
monopole--anti-monopole pair (connected by a $Z$--string) in an
unstable saddle point bound state.  This provides an example for the
monopole--pair unbinding mechanism expected to work at the
electroweak phase transition.
The definitions of defects developed here will be used in
future studies of topological aspects of this
transition.
}

\newpage
\baselineskip=14pt
%%% For alphabetic footnotes indices in text  %%%%
\setcounter{footnote}{0}
\renewcommand{\thefootnote}{\alph{footnote}}
%%%%%%%%%%%%%%%%%%%%%%%%%%%%%%%%%%%%%%%%%%%%%%%%%%

\section{Introduction}

It might be surprising that topological aspects of the electroweak
theory (although being under discussion already for some time
outside the lattice community)
have not paid due attention to by people doing
lattice simulations of the electroweak phase transition.
With the present paper, we are entering
investigations in that direction for the
standard
$SU(2)$ Higgs model. The phase transition of this model
(without supersymmetric extensions) has lost most
of its phenomenological
appeal as a viable scenario explaining the
generation of baryon asymmetry (taking the present lower limit of the Higgs
mass into account).
From the non--perturbative point of view in general,
this model remains attractive, however,
as a laboratory
for studying the strong coupling features of high temperature
gauge field theory coupled to matter.

In the case of QCD,
in contrast to the situation in electroweak theory,
the thermal transition between the hadronic and the
quark--gluon plasma phase is under intensive study
with respect to its topological aspects.
The transition is known to be
accompanied by a restructuring of the Euclidean field
configurations concerning their instanton and monopole
content, the latter, however, being detected only by choosing particular gauges.
The most promising one in the case of QCD seems to be the maximally Abelian
gauge.
The Abelian degrees
of freedom in this gauge~\cite{KrLaScWi87}
were shown to be relevant for various dynamical properties of the
confining phase of
Yang--Mills theories realizing the dual superconductor scenario of
confinement~\cite{tH81}.
This development can be followed in Refs.~\cite{Rev's}.
The Abelian degrees of freedom provide the dominant contribution
to the non-Abelian string tension in the $SU(2)$
gluodynamics~\cite{SuzukiBali}.

We are wondering whether Abelian monopoles, perhaps
in a particular Abelian projection, may also play a
non--trivial role in the dynamics of the electroweak theory at high
temperature.
One outstanding feature of the symmetric phase of the $SU(2)$ fundamental
Higgs model is its $3D$ (magnetic) confinement property. Of course,
this is also captured in the dimensionally reduced variant of the model
which is able to describe the thermal transition with high quantitative
accuracy~\cite{Thermal}.
With respect to $3D$ confinement,
the symmetric phase resembles very much
the $3D$ pure Yang-Mills theory investigated by Teper~\cite{Teper}.
Some years ago, Bornyakov and Grygoryev~\cite{Bornyakov}
attempted to identify the agents of confinement
in $3D$ pure gauge theory by applying the Abelian projection technique
to this model. They found that Abelian monopoles occur with a density
becoming constant in the continuum limit $\beta \to \infty$ measured in
natural units $g_3^2=g_4^2~T$. It is known that dimensional reduction cannot
describe the nature of the deconfinement transition in QCD. In the
electroweak case, however, dimensional reduction has been shown to work
very well in the temperature range of interest.

Therefore it seems natural to start our consideration of topological
restructuring at the electroweak phase transition by choosing the
$3D$ formulation and applying the Abelian projection approach. We
will compare the maximally Abelian gauge with a gauge independent
prescription which corresponds to Abelian
projection from the unitary gauge (for the Higgs field in the
fundamental representation).  The first approach requires an iterative
procedure, and usually the monopole content suffers from gauge
dependencies.  The second one does not need any gauge fixing.

Instead of addressing directly the phase transition, we will test our
tools analyzing certain {\it classical} field configurations that exist
in the broken phase and are believed to contain pairs of non-Abelian
Nambu monopoles~\cite{Ma83,Na77} (non--Abelian monopolium).
Numerical work is in progress~\cite{InProgress}
indicating that the density of the corresponding type of monopoles
to be defined in the present paper indeed
behaves almost as an
disorder parameter characterizing the
$3D$ confining, symmetric phase in the dimensionally reduced
theory.  The condensation of vortices, presumably the correct
order parameter, is presently under study. But if
monopole--anti-monopole pairs become unbound in the symmetric phase,
being invisibly bound in the broken phase, couldn't be the monopole
separation inside the sphalerons the precursor of this mechanism ?

The sphaleron~\cite{DaHaNe74,Ma83,KlMa84} is believed to be important
for the relaxation of an eventual baryon number asymmetry after the
electroweak phase transition is completed~\cite{SphRev}. In the
standard electroweak model all constraints concerning the strength of
the electroweak phase transition are derived from the requirement
that the thermal barrier factor $exp(-E_{sphal}(T_c)/T_c)$ should be
sufficiently small to suppress the washing--out of the baryon number
in the broken, lower--temperature phase. In the present work we are
only interested to learn how the electroweak sphaleron looks like in
various Abelian projections. We study the electroweak sphaleron using
the $SU(2)$ Higgs model since due to the smallness of the
Weinberg angle $\theta_W$ the $U(1)$ component of the electroweak
group $SU(2) \times U(1)$ has little effect on the sphaleron
properties and also the influence of the standard model fermions
on the sphaleron solution is quite small. Thus the properties of the
sphaleron in the standard electroweak model are basically
determined by the $SU(2)$ Higgs sector~\cite{SphRev}. The $3D$
sphaleron is perfectly known on the lattice due to the work of
Garcia Perez and van Baal~\cite{PeBa95}.  They used a $3D$ variant
of the $SU(2)$ fundamental Higgs theory in order to construct
sphalerons as saddle point solutions of the lattice energy
functional~\cite{PeBa95}.

In the maximally Abelian gauge we find
that the sphaleron contains a highly symmetric structure of Abelian
monopole--anti-monopole pairs connected by vortex strings.
We also study the other definition of defects which corresponds to Abelian
projection from the unitary gauge. The sphaleron configurations have
been generated just in this gauge~\cite{PeBa95}, but the prescription
works without any gauge fixing.  We show, that in {\it this} Abelian
projection the sphaleron contains a Nambu monopole--anti-monopole
pair~\cite{Na77} connected by a $Z$-vortex string~\cite{Ma83,Na77}
(non--Abelian monopolium\footnote{One of the Abelian monopole pairs
found in the maximally Abelian gauge {\it is} the Nambu monopole
pair.}). This result comes not unexpectedly in view of some
investigations~\cite{BaVaBu94,Hi94} done in the continuum.

The structure of the paper is as follows. In
Section~2 we formulate the maximally Abelian projection of the
$SU(2)$ fundamental Higgs model with emphasis on the Higgs degrees of
freedom. We show that this model in the maximally Abelian projection
contains Abelian monopoles and two types of
Abrikosov--Nielsen--Olesen strings~\cite{ANO}. In Section~3 we
present our gauge invariant lattice definitions of the Nambu monopole
and the $Z$--string.  Section~4 is devoted to the analysis of some
$3D$ sphaleron configurations (produced by~\cite{PeBa95}) and
Section~5 to the interpretation of our findings.

\section{The maximally Abelian Projection of $SU(2)$--Higgs Theory}

The maximally Abelian gauge is defined~\cite{KrLaScWi87} to maximize some
gauge--noninvariant functional $R[U]$ by suitable gauge transformations,
where $R[U] =
\sum_l {\rm Tr} \, (U_l \sigma^3 U^+_l \sigma^3)$. $U_l$ denotes
a link representing a
$SU(2)$ gauge field and $\sigma^3$ is one of the Pauli matrices. The
functional $R$ is still invariant under $U(1)$ gauge transformations,
$\Omega^{abel}_x = e^{i \sigma_3 \, \alpha_x}$, $\alpha_x \in [0,2
\pi)$. The gauge condition fixes the $SU(2)$ gauge freedom up
to the $U(1)$ subgroup.

One speaks about Abelian projection if
Abelian link phases $\theta_l$ are
extracted from the diagonal elements of the $SU(2)$ gauge field $U$ according to
$\theta_l = \arg U^{11}_l \in [-\pi,\pi)$.
Usually, the Abelian projection is done after the maximally Abelian
gauge has been chosen.
Under the residual $U(1)$
gauge transformations the field $\theta_l$ behaves as an Abelian gauge
field: $\theta_{x,\mu} \to \theta_{x,\mu} + \alpha_{x+\hat\mu} -
\alpha_x$ mod $2\pi$. The components of the $SU(2)$--Higgs field
$\Phi={(\phi^{(1)},\phi^{(2)})}^T$ transform as follows: $\phi^{(1)}
\to e^{i \alpha} \phi^{(1)}$ and $\phi^{(2)} \to e^{-i \alpha}
\phi^{(2)}$.  Thus the fields $\phi^{(1)}$ and $\phi^{(2)}$ carry
Abelian charges $+1$ and $-1$, respectively.

Therefore, the $SU(2)$--Higgs theory in the Abelian projection can be
considered as a theory which contains a compact Abelian gauge field
$\theta_l$ and two charged Abelian $scalar$ fields\footnote{The
fields put in the Abelian gauge comprise also the non--diagonal
components of the $W$ fields which behave as Abelian matter $vector$
fields. We do not pay special attention to these non--diagonal gauge
field components in this paper.} $\phi^{(1)}$ and $\phi^{(2)}$. The
reduced theory possesses two types of topological defects, Abelian
monopoles (due to the compactness of the residual Abelian group) and
Abelian vortices (due to the presence of the charged scalar fields).

The Abelian plaquette $\theta_{x,\mu\nu} = \theta_{x,\mu} +
\theta_{x+\hat\mu,\nu} - \theta_{x+\hat\nu,\mu} - \theta_{x,\nu}$ can
be decomposed into two parts: $\theta_{x,\mu\nu} = {\bar
\theta}_{x,\mu\nu} + 2 \pi m_{x,\mu\nu}$. Here ${\bar
\theta}_{x,\mu\nu} \in [-\pi,\pi)$ is the electromagnetic flux
through the plaquette $P_{x,\mu\nu}$ and $m_{x,\mu\nu}$ is an
integer associated with a Dirac string. The Abelian monopole
charge inside a cube $C$ is identified as follows~\cite{DGT}:
\beqn
j_C = {(\dd m)}_P\equiv \sum_{P \in \partial C} m_P\,,
\label{j_Abelian}
\eeqn
where the summation is taken over the plaquettes which form the
boundary of the cube~$C$ and $\dd$ denotes the lattice differential.

Actually, the effective Abelian theory
possesses {\it two} types of Abelian vortices since
there are two Abelian charged fields. The vorticity numbers (of sort $i$)
$\sigma^{(i)}$ carried by the plaquette $P_{x,\mu\nu}$ are
given by the following equations~\cite{ChPoZu94}:
\beqn
\sigma^{(i)}_{x,\mu\nu} = m_{x,\mu\nu} -
l^{(i)}_{x,\mu\nu} \,, \quad
l^{(i)}_{x,\mu\nu} = l^{(i)}_{x,\mu}
+ l^{(i)}_{x+\hat\mu,\nu} - l^{(i)}_{x+\hat\nu,\mu} -
l^{(i)}_{x,\nu}\,,\quad i=1,2\,,
\label{SigmaAbelian}
\eeqn
where the integer-valued link variables $l^{(i)}$ are defined, in terms of the
link angles $\theta_{x,\mu}$ and the phases of the respective
(upper or lower) Higgs field components $\varphi^{(i)}_x = \arg \phi^{(i)}_x$,
through the usual decomposition
\beqn
\mp \varphi^{(i)}_x + \theta_{x,\mu} \pm \varphi^{(i)}_{x+\hat\mu}
- 2 \pi l^{(i)}_{x,\mu} \in [-\pi,\pi)\,,
\qquad
 i=1,2\,.
\eeqn

The vorticity number $\sigma^{(i)}_P$ is equal to the number of
type-$i$ vortices
penetrating the plaquette $P$. The vortex trajectories are defined as
the set of oriented links which are dual to the plaquettes with
non--zero vorticity number. On can check that the Abelian
vortices of both types end on the Abelian monopoles, {\it i. e.}
$\dd
\sigma^{(i)} = j$.

\section{The Nambu Monopoles and the $Z$--Strings}

There is a gauge invariant and quantized lattice definition of
another type of magnetic excitations of the $SU(2)$ Higgs theory, the
Nambu monopole~\cite{Na77}. We define a composite adjoint unit vector
field $n^a_x$ by
\beqn
n^a_x =
- \frac{\left(\Phi^+_x, \sigma^a \Phi_x\right)} {\left(\Phi^+_x,
\Phi_x\right)}\,, \qquad n_x = n^a_x \, \sigma^a\,.
\label{n}
\eeqn
In the following definition of the Nambu monopole the field $n_x$
plays a role similar to the direction of the adjoint Higgs field in
the definition of the 't~Hooft--Polyakov monopole~\cite{tHPo74} in
the Georgi--Glashow model\footnote{For a discussion and application of
various definitions of a
magnetic charge to investigate this model we refer to
Ref.~\cite{bornyakov1}.}.

The most transparent definition of the Nambu monopole can be given in
the unitary gauge $\Phi={(0,\phi)}^T$ ($n^a \equiv \delta^{a3}$), where
$\phi$ is some complex--valued scalar. This gauge condition leaves
Abelian transformations $\Omega^u_x = e^{-i \sigma^3 \alpha_x}$ free,
with $\alpha_x \in [0,2\,\pi)$. The
superscript $^u$ refers to the
unitary gauge. The phase $\theta^u_l= \arg U^{11}_l$
of some link behaves as a
compact $U(1)$ gauge field with respect to the residual Abelian gauge
group: $\theta^u_{x,\mu} \to \theta^u_{x,\mu} + \alpha_{x+\hat\mu} -
\alpha_x\,\, {\rm mod}\, 2 \pi$.
In the continuum $SU(2)$ Higgs theory~\cite{Na77} the
$Z$--magnetic flux coincides in the unitary gauge with the Abelian
magnetic flux\footnote{Note that in the standard electroweak model
(at non-zero Weinberg angle $\theta_W$)
the $Z$--flux acquires also a contribution from the $U(1)$ sector.
We do not discuss this case in the present paper.} of the field
$\theta^u_l$.  Therefore, in the unitary gauge the Nambu monopoles
can be identified with the Abelian magnetic defects in the field
$\theta^u_l$.

Now the usual DeGrand--Toussaint construction~\cite{DGT}
can be applied to the
field $\theta^u_l$ in order to define the
$Z$--charge of the Nambu monopole inside a three-dimensional
cube~$C$:
\beqn
j^u_C = \sum_{P \in \partial C} m^u_P \equiv
- \frac{1}{2\pi} \sum_{P \in \partial C} {\bar \theta}^u_P\,, \quad
{\bar \theta}^u_{x,\mu\nu} =
\theta^u_{x,\mu\nu} - 2 \pi m^u_{x,\mu\nu} \in [-\pi,\pi)\,,
\label{j_N}
\eeqn
where the summation is taken over the plaquettes which form the
boundary of the cube~$C$.

There is, however, a gauge invariant way to define
the flux ${\bar \theta}^u_P$ which proceeds as follows.
First a new set of links $V_l$
depending on $U_l$  and $n_x$
is introduced
by the  following construction
\beqn
V_{x,\mu}(U,n) = U_{x,\mu} + n_x U_{x,\mu}
n_{x + \hat \mu}\,.
\label{V}
\eeqn
Under general gauge transformations, $V_l$ links transform like $U_l$ links.
They intertwine the adjoint field $n_x$ between neighboring places,
\beqn
n_x V_{x,\mu}=V_{x,\mu} n_{x+\hat\mu}.
\label{intertwine}
\eeqn
Before the fluxes of the $V_l$'s are evaluated the new links
have to be normalized to $SU(2)$ giving
\beqn
A_{x,\mu}(U,n) = \frac{V_{x,\mu}(U,n)}{
\sqrt{\frac{1}{2} {\rm Tr} \Bigl[V^+_{x,\mu}(U,n)\, V_{x,\mu}(U,n)
\Bigr]}}\,.
\label{A}
\eeqn
The gauge invariant
flux ${\bar \theta}^u_P$ is now calculable as
\beqn
{\bar \theta}^u_{x,\mu\nu} (U,n) = \arg \Bigl( {\bf Tr}
 \left[(\bbbone + n_x) A_{x,\mu} A_{x +\hat\mu,\nu}
 A^+_{x + \hat\nu,\mu} A^+_{x,\nu} \right]\Bigr)\,.
 \label{AP}
\eeqn
The Abelian plaquette \eq{AP} is a gauge invariant object because the
field $A_l$ transforms as an $SU(2)$ link field and the vector $n^a_x$
as an adjoint matter field. In the unitary gauge, when $n \equiv
\sigma^3$, the field $A_l$ is exactly diagonal
\beqn
A_{x,\mu}(U,\sigma^3) = {\rm diag}\, (e^{i \theta^u_{x,\mu}}\,,\,
e^{ - i \theta^u_{x,\mu}})\,,\qquad \theta^u_{x,\mu} =
{\rm arg} U^{11}_{x,\mu}
\label{theta_abel}
\eeqn
due to cancellations in (\ref{V}). Because of (\ref{intertwine}) the
Abelian plaquette is independent of which corner is chosen in order
to project the non--Abelian plaquette
onto $n_x$.
The formulae (\ref{n}-\ref{AP}) give the gauge--invariant lattice
definition of the Nambu monopole.

The lattice $Z$--string is primarily defined in the unitary gauge,
$\Phi={(0,\phi)}^T$. Under the residual Abelian gauge transformations
the field $\phi_x$ behaves as follows: $\phi_x \to e^{i \alpha_x} \phi_x$.
Therefore in the unitary gauge the lower component $\phi_x$ of the
doublet Higgs field $\Phi_x$ has unit electric charge with respect to the
Abelian gauge field $\theta^u_l$.
The $Z$--string~\cite{Ma83,Na77} can be
considered as the
Abrikosov--Nielsen--Olesen
vortex solution~\cite{ANO} embedded~\cite{VaBa69,BaVaBu94} into the
electroweak theory. As long as we are in the unitary gauge the $Z$--strings
can be detected as the vortex topological defects in the Abelian
matter field~$\phi_x$. Then the $Z$-vorticity number through
the plaquette $P_{x,\mu\nu}$
can be defined as follows:
\beqn
\sigma^u_{x,\mu\nu} = m^u_{x,\mu\nu} - l^u_{x,\mu\nu}
\equiv -\frac{1}{2\pi} \Bigl(
{\bar \theta}^u_{x,\mu\nu} - \chi^u_{x,\mu\nu}\Bigr) \,,
\label{SigmaN}
\eeqn
where the link variables $l^u_l$ and $\chi^u_l$
are the result of the decomposition
\beqn
\chi^u_{x,\mu} =
\varphi^u_x + \theta^u_{x,\mu} - \varphi^u_{x+\hat\mu} - 2 \pi
l^u_{x,\mu} \in [-\pi,\pi)\,, \qquad \varphi^u_x = \arg \phi_x\,.
\eeqn
Finally, the non--integer part is used to evaluate the
plaquette field $\chi_{x,\mu\nu} \equiv {(\dd \chi)}_{x,\mu\nu}$. But there
is an alternative, gauge independent way to define the field
$\chi^u_l$ as follows:
\beqn
 \chi^u_{x,\mu} = - \arg\left(\Phi^+_x, A_{x,\mu}
 \Phi_{x + \hat\mu}\right)\,, \label{chiG}
\eeqn
where the link field $A_l$ is defined in \eq{A}.
$Z$--vortices begin and end on
Nambu (anti-)mo\-no\-po\-les: $\dd \sigma^u = j^u$.
Equations
(\ref{SigmaN}-\ref{chiG}) comprehend the gauge invariant
lattice definition of the $Z$--vortex.

\section{Topological Content of Sphaleron:
A Few Exercises with Lattices Sphaleron
Configurations}

The electroweak sphaleron (at Weinberg angle $\theta_W=0$) is a saddle point
solution of the static equations of motion of the $SU(2)$--Higgs
theory. We have studied four three--dimensional electroweak sphaleron
configurations which have been originally obtained in
Refs.~\cite{PeBa95} on a $16^3$ lattice (with periodic boundary
conditions).  The sphalerons were prepared by a saddle point cooling
algorithm, {\it i.e.} cooling with respect to an action which
expresses the square of the equation of motion (as written on the
lattice). For details, one should consult Refs.~\cite{PeBa95}. The
four sphaleron configurations differ from each other by the physical
size of the lattice, $L M_W=4$ and $L M_W=4.8$, and by different
Higgs mass parameters of the model, $M_H=M_W$ and $M_H=0.75~M_W$. The
configurations have been kindly provided to us by the authors of
Refs.~\cite{PeBa95}.

First, we observe that the sphaleron configurations happen to be
Abelian to a high degree in the (quasi--unitary) gauge they are produced in,
$\Phi={(\phi,0)}^T$ with $\phi_x$ real--valued.
The volume average of the sum of the squared
diagonal elements in the $SU(2)$ link matrices $U_l$ is not less
than $0.96$. If the Abelian link angles are extracted in this
{\it not yet maximally Abelian} gauge,
the
Nambu $M\overline{M}$ structure hidden in the core of the sphaleron
can immediately made visible in the appearance of Abelian
monopoles\footnote{Note, that the definition of the Nambu
monopole (\ref{n}-\ref{AP}) is the same both in quasi-unitary and in
unitary gauges due to the $n \to - n$ invariance of the field \eq{A}.}.
Nambu monopoles {\it are} Abelian monopoles in the (quasi-) unitary gauge.
All four sphalerons have
the Nambu monopole $M$ and the Nambu anti-monopole $\overline{M}$ at
a distance of two lattice spacings at the same place in the
lattice. This similarity can be explained by the fact that all sphaleron
solutions are derived from a single one by adapting the parameters
for the new saddle-point cooling~\cite{private}. We relate our
observation to the result obtained in the framework of continuum
field theory~\cite{BaVaBu94,Hi94} that the sphaleron should contain a
pair of Nambu monopole and anti--monopole connected by a piece of
$Z$--vortex.

The same picture is reproduced by the gauge invariant
measuring routine outlined above.
The configuration is visualized in Fig.~1.  The big points denote
the Nambu monopole and anti-monopole, the line in between is the
$Z$--vortex trajectory and the volume occupied by the sphaleron is
marked by the cloud of small points.  Their  density is inversely
proportional to the modulus of the scalar field.  Regions where the
length of the scalar field is bigger than $0.75$ are not shown.

After looking into each sphaleron in the quasi--unitary gauge
$\Phi={(\phi,0)}^T$ with $\phi_x$ real--valued, we attempted
to put them into the maximally Abelian gauge.
This has been done with the standard
algorithm adapted to $3D$. There is always an obstruction to reach full
Abelianicity. We have searched for the maximum of the gauge-fixing
functional $R[U]$
over 100 random gauge copies
(to control possible Gribov copies)
of each original sphaleron
configuration.
Our stopping criterion for the gauge cooling iterations was that
we continue to cool if
the {\it volume minimum} of the trace of the local gauge
transformation, $1\slash 2 \, {\rm Tr} g_x$ is less than
$1-10^{-8}$.  It is remarkable that,
for each Gribov copy, at the end of the gauge cooling always one of the
Abelian monopole pairs we detected was identical with
the Nambu $M\overline{M}$ pair. As the gauge cooling proceeds,
few additional pairs of Abelian monopoles pop up and disappear while the
Nambu monopole pair always remains among the Abelian monopoles.

Finally put into the maximally Abelian gauge, all investigated
sphaleron configurations have Abelian
monopoles forming a
rotationally invariant (under the cubic group) configuration in the
very center of the sphaleron.

For the Higgs mass being equal to the $W$ mass the
Abelian monopolium structure is shown in Fig.~2.
Now, the big points denote the
Abelian monopoles and anti-monopoles and the lines are Abelian
vortex trajectories.  Figs.~2(a,b) represent the configuration
enclosed in the
smaller lattice volume. In particular, Fig.~2(a) shows the type-1
vortices and Fig.~2(b)
the type-2 vortices. Similarly, in Figs.~2(c,d) we visualize the
sphaleron that has been created in the physically larger volume. At
the chosen mass ratio there is apparently no volume dependence.

The sphalerons with the mass ratio $M_H/M_W=0.75$ look somewhat different
as can be seen in Fig.~3.
In Figs.~3(a,b) the configuration enclosed in the
smaller lattice volume is depicted, one time showing the type-1 and
the other time the type-2 vortices. Analogously, Figs.~3(c,d)
allow to have a look into
the sphaleron in the larger lattice volume. Again, the lattice volume
plays no important role for the structure of the core.
The comparison with Fig~2. suggests that there is an effect of the
Higgs mass on the distribution of vortices.
In the case of higher Higgs mass there are
long vortex trajectories
sweeping out in a random walk through the region of high energy density.
No such long vortex trajectories are detected in Fig.~3.

\section{Conclusion}

The fundamental $SU(2)$--Higgs model in $3D$ can be topologically
analyzed in terms of {\it Abelian monopoles},  for instance in the
symmetric phase where these should be related to the $3D$--confining
properties.  The dimensionally reduced high temperature Higgs theory
has a string tension  of approximately the same strength as the pure
$3D$ gauge theory.  The Abelian reduction of the Higgs model from the
maximally Abelian gauge leaves two Abelian Higgs fields (with charge
$1$ and $-1$ with respect to the $U(1)$ subgroup).  Correspondingly,
there exist two types of {\it Abelian vortices}, which are closed or
connect a monopole with an anti--monopole.

In this paper gauge independent lattice definitions of {\it
non--Abelian monopoles and vortices} (the Nambu monopoles and the
$Z$--vortices) have been formulated which describe embedded,
topologically unstable defects within the fundamental Higgs model.
The Nambu monopole current is topologically conserved in $4D$. The
new kind of monopole is identical with the normal Abelian monopole
iff the Abelian projection (which leads to the latter) is done in the unitary
gauge, but they are still strongly correlated if (as usual) the maximally
Abelian
gauge is chosen.

There are ideas expressed in the literature~\cite{BaVaBu94,Hi94} that
the sphaleron saddle point configuration, from the $3D$ point of
view, should consists of a pair of Nambu monopoles stretched along a
$Z$-vortex string.  We have analyzed a few
lattice sphaleron configurations looking from the two perspectives
explained above, in terms of Abelian monopoles and vortices on one
hand and of non--Abelian Nambu monopoles and $Z$--vortices on the
other.  This analysis has covered Higgs masses $M_H/M_W= 1.0$ and $0.75$ and
volumes ranging from $V M_W^3=4.0^3$ to $4.8^3$ on a $16^3$ lattice.
Since the lattice saddle point configurations have been provided in
the quasi--unitary gauge $\Phi={(\phi,0)}^T$ with $\phi_x$
real--valued\footnote{This is the gauge chosen for running the
extremization in Ref.~\cite{PeBa95}.},
the two pictures are identical for the original
sphalerons.  There is always just one Nambu monopolium in the center
of the sphaleron separated by a distance $d=2a$.

In fact, the original configurations proved to be already relatively
Abelian, but then gauge cooling has been applied to put
them into the maximally Abelian gauge.
In this gauge an Abelian multi--(anti)monopole
configuration appears which is maximally
symmetric under the cubic rotation group.
One of the Abelian monopole--anti-monopole
pairs is identical with the non--Abelian Nambu $M\overline{M}$ pair.
As the result of gauge cooling,
this structure is found
for all
$O(100)$ random gauge copies
per sphalerons that we have
prepared to start from.
The exact trajectories of the Abelian vortices in the 'maximally' Abelian
gauge differs from copy to copy.
Mostly they are of length $2a$.
Our results are indicative for an effect that the Higgs mass has
on the distribution of
the Abelian vortices. In the case of higher Higgs mass
there is always one vortex (either of type 1 or type 2) of extension
(length) greater than $2a$.

Simulations are now under way in order to clarify the dynamical role
of the topological defects discussed here in the context of the
thermal Higgs phase transition~\cite{InProgress}.

\section*{Acknowledgements}

We are very grateful to Pierre van Baal and Margarita Garcia--Perez 
for providing the sphaleron lattice configurations. The authors have 
be\-ne\-fited from dis\-cus\-sions with M.~M\"ul\-ler--Preussker, 
M.~I.~Polikarpov and T.~Suzuki. We thank A.~Schiller for pointing out 
the misprint in eq.\eq{chiG} and also for collaboration on a related 
project.

M.~N.~Ch. acknowledges the kind hospitality of the Department of
Theoretical Physics of Humboldt University (Berlin) and Kanazawa
University. M.~N.~Ch. was supported by the DFG grant 436 RUS 113/29/23.
M.~N.~Ch. and F.~V.~G. were partially supported by the JSPS Program on
Japan--FSU scientists collaboration, and also by the grants INTAS-94-0840,
INTAS--RFBR-95-0681 and the grant No. 96-02-17230a of the Russian
Foundation for Fundamental Sciences. E.--M.~Ilgenfritz was supported by DFG
under grant Mu932/1-4.

%\newpage

\newpage

\begin{figure*}[t]
\begin{center}
\hspace{-0.8cm}\epsfxsize=13.4cm\epsffile{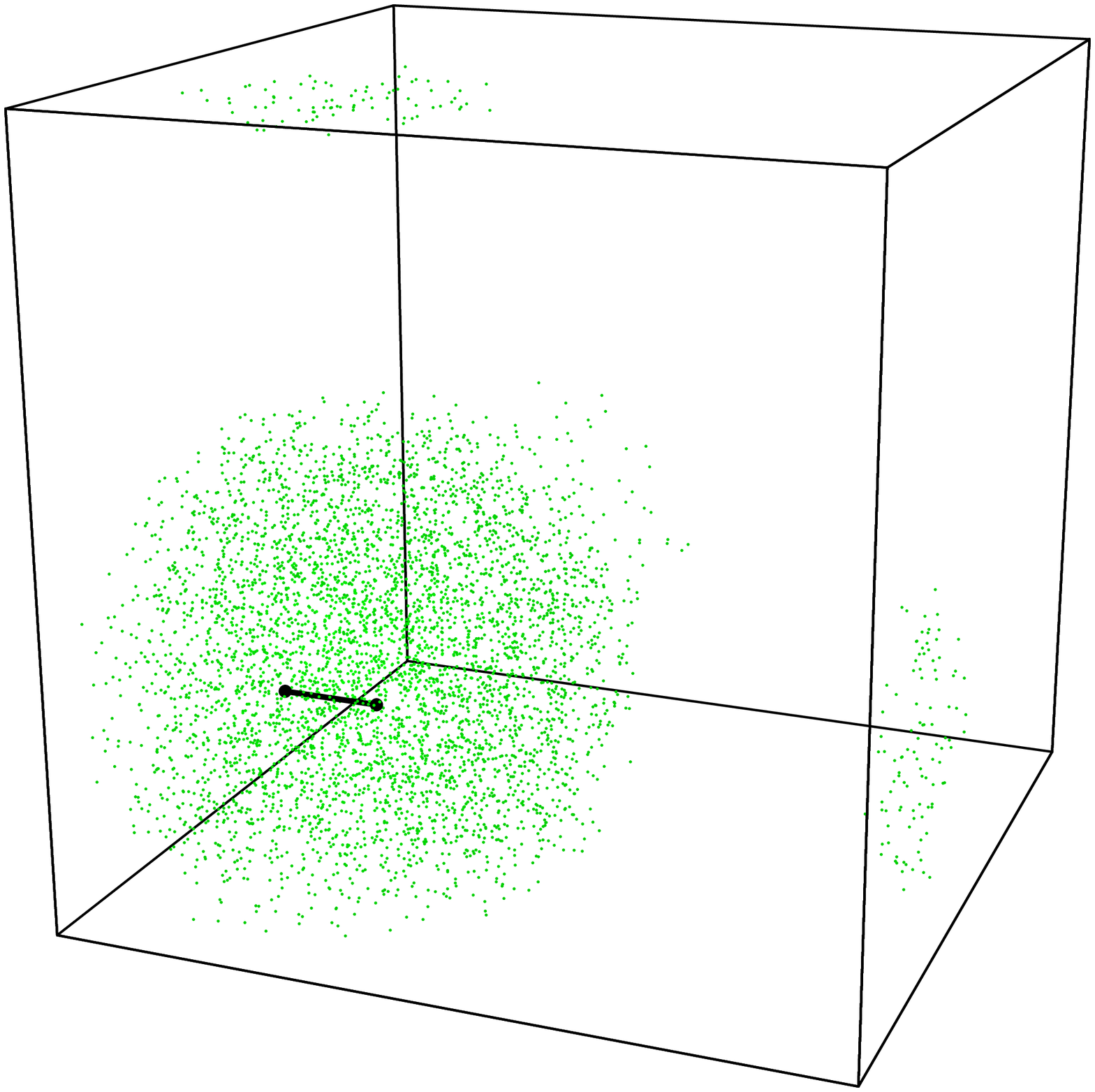} \\
\end{center}
\vspace{-0.5cm}
\caption{Nambu monopole-anti-monopole pair with Z-vortex in between
(non-Abelian monopolium) inside the lattice sphaleron. The
positions are the same for all four sphaleron configurations.}
\end{figure*}

\begin{figure*}[t]
\begin{center}
\begin{tabular}{cc}
\hspace{-0.8cm}\epsfxsize=6.7cm\epsffile{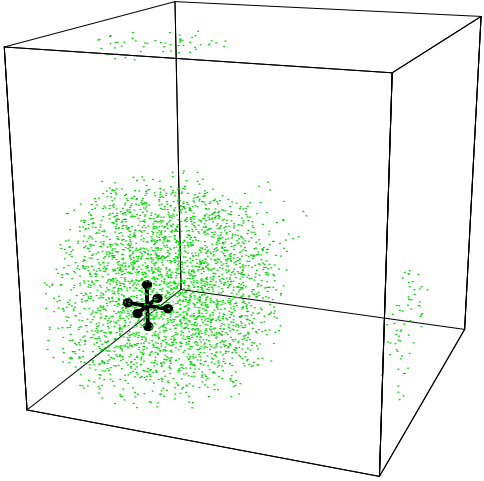} &
\hspace{0.8cm}\epsfxsize=6.7cm\epsffile{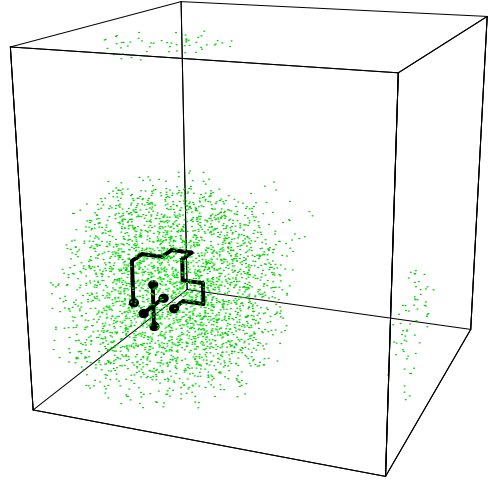} \\
(a) & (b)   \vspace{-0.3cm} \\ \\
\hspace{-0.8cm}\epsfxsize=6.7cm\epsffile{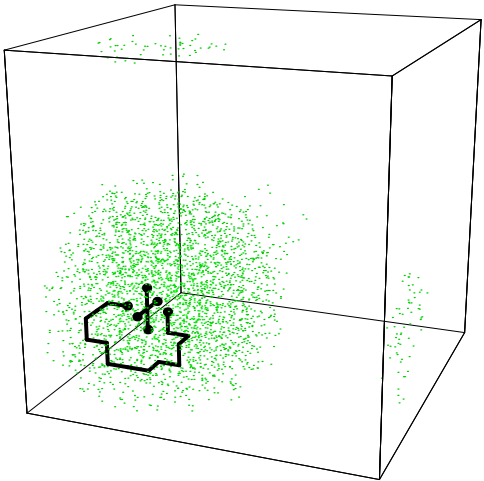} &
\hspace{0.8cm}\epsfxsize=6.7cm\epsffile{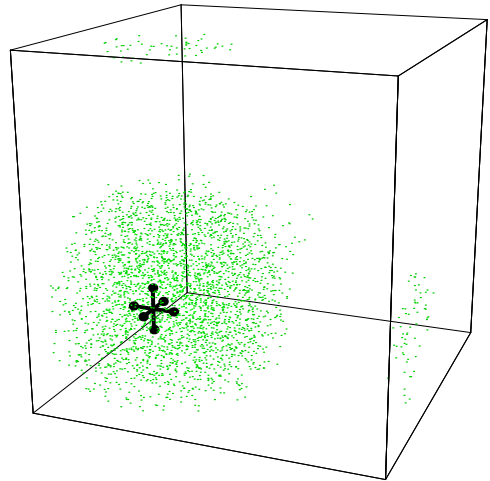} \\
(c) & (d)   \vspace{-0.3cm} \\ \\
\end{tabular}
\end{center}
\vspace{-0.5cm}
\caption{The sphalerons with $M_H=M_W$ in the volumes $L M_W=4$
(above) and $L M_W=4.8$ (below) in the Maximal Abelian gauge with
their Abelian monopole content and with Abelian vortices of the type
I (left) and the type II (right).}
\end{figure*}

\begin{figure*}[t]
\begin{center}
\begin{tabular}{cc}
\hspace{-0.8cm}\epsfxsize=6.7cm\epsffile{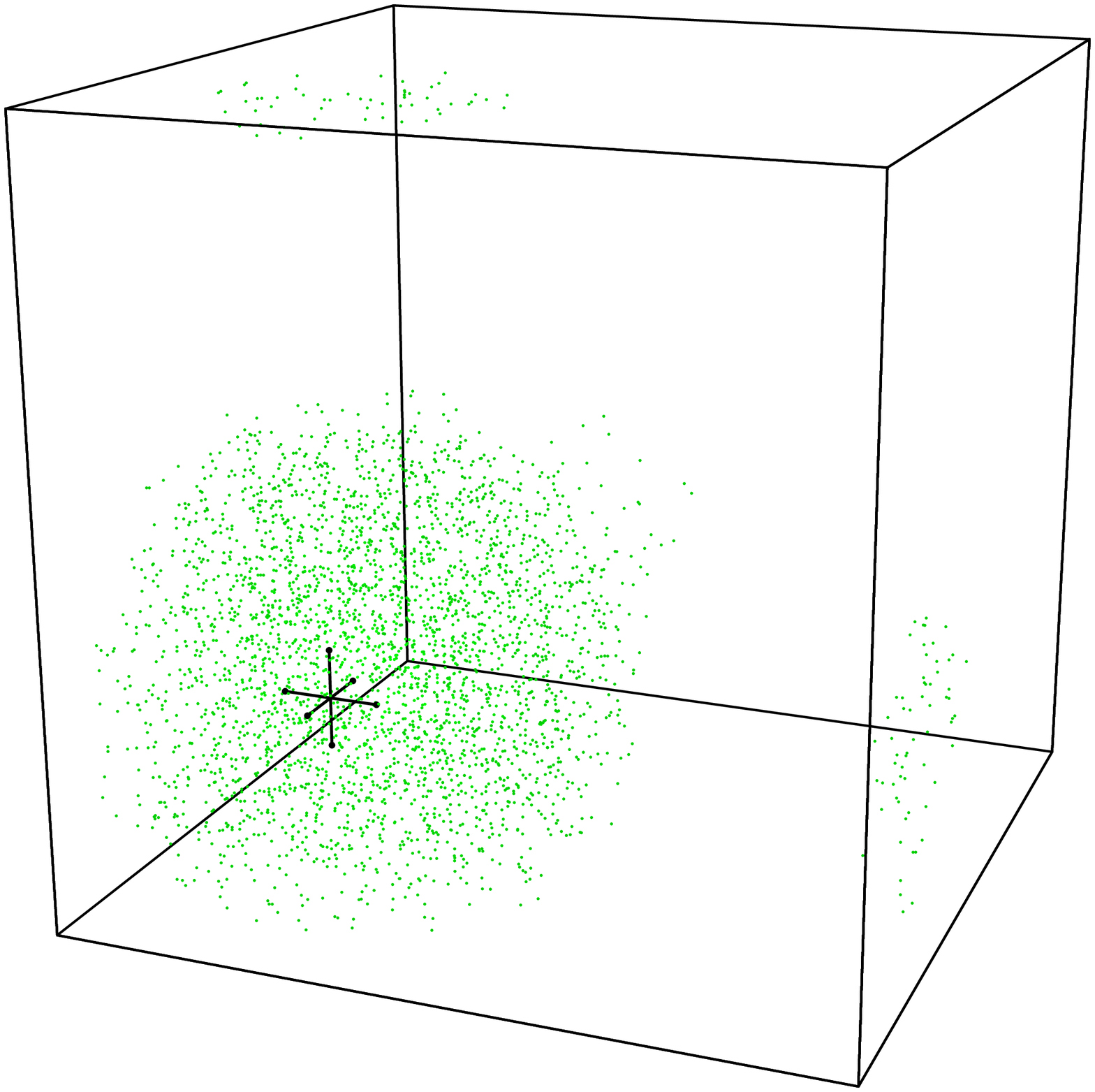} &
\hspace{0.8cm}\epsfxsize=6.7cm\epsffile{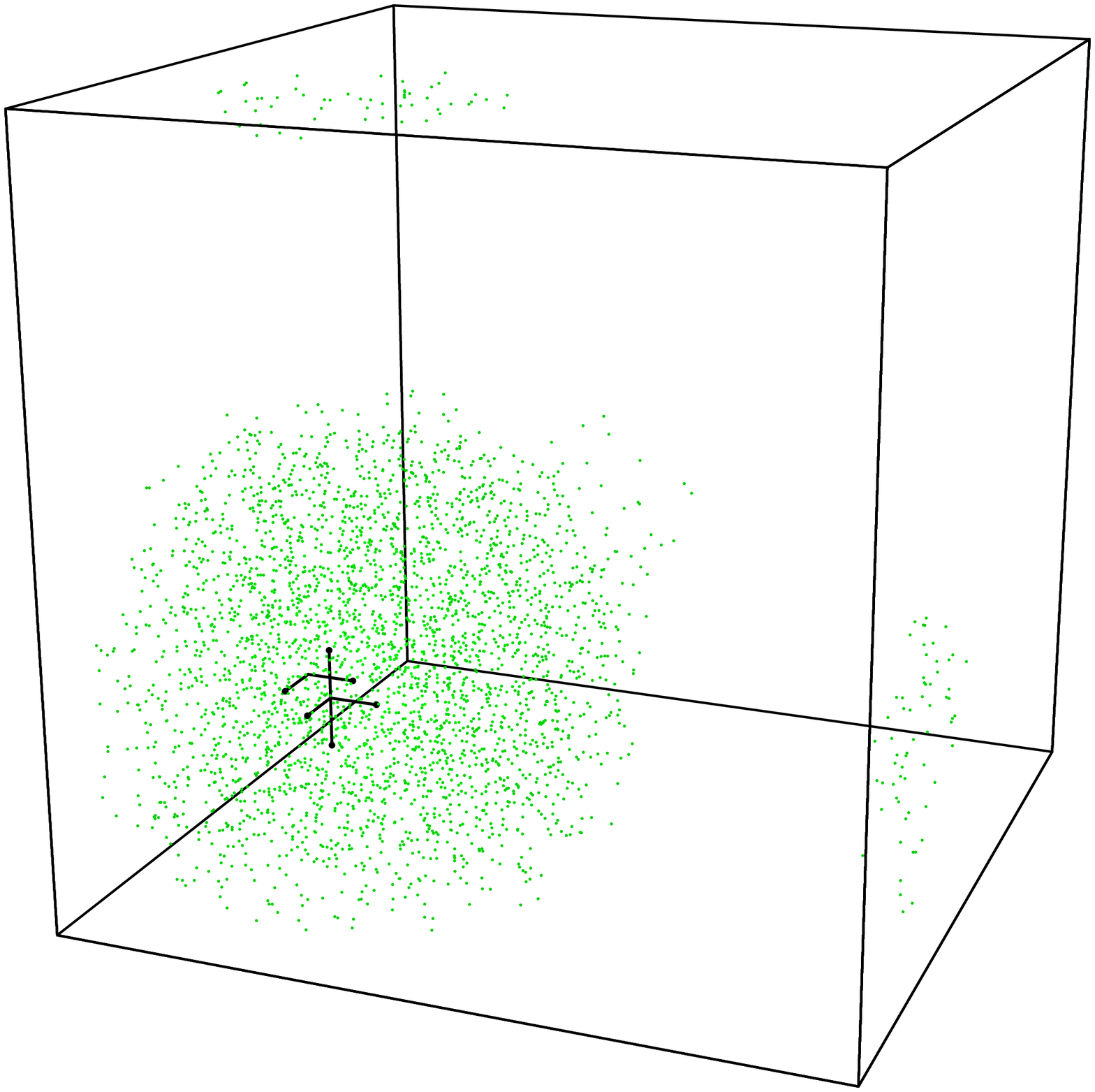} \\
(a) & (b)   \vspace{-0.3cm} \\ \\
\hspace{-0.8cm}\epsfxsize=6.7cm\epsffile{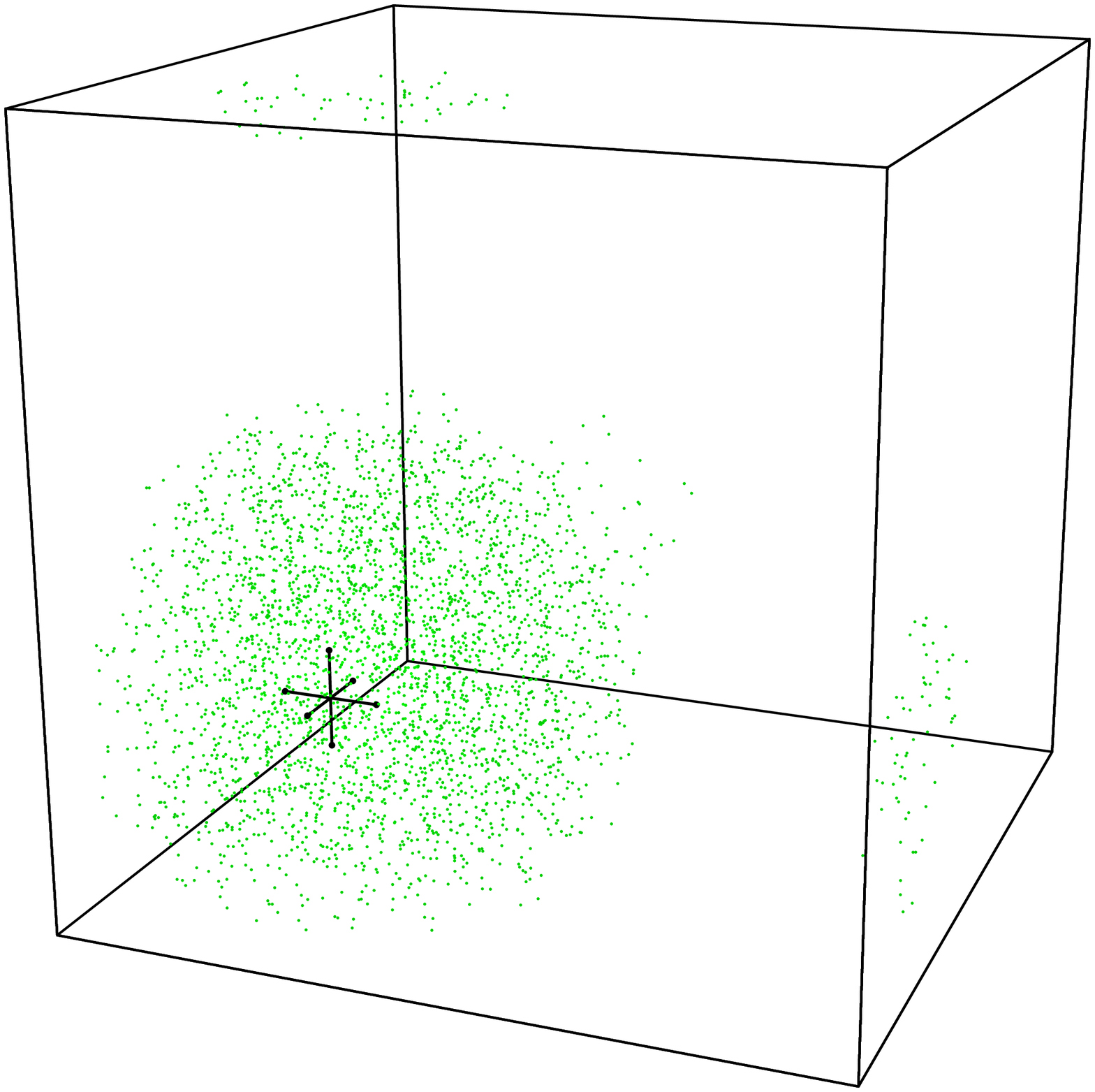} &
\hspace{0.8cm}\epsfxsize=6.7cm\epsffile{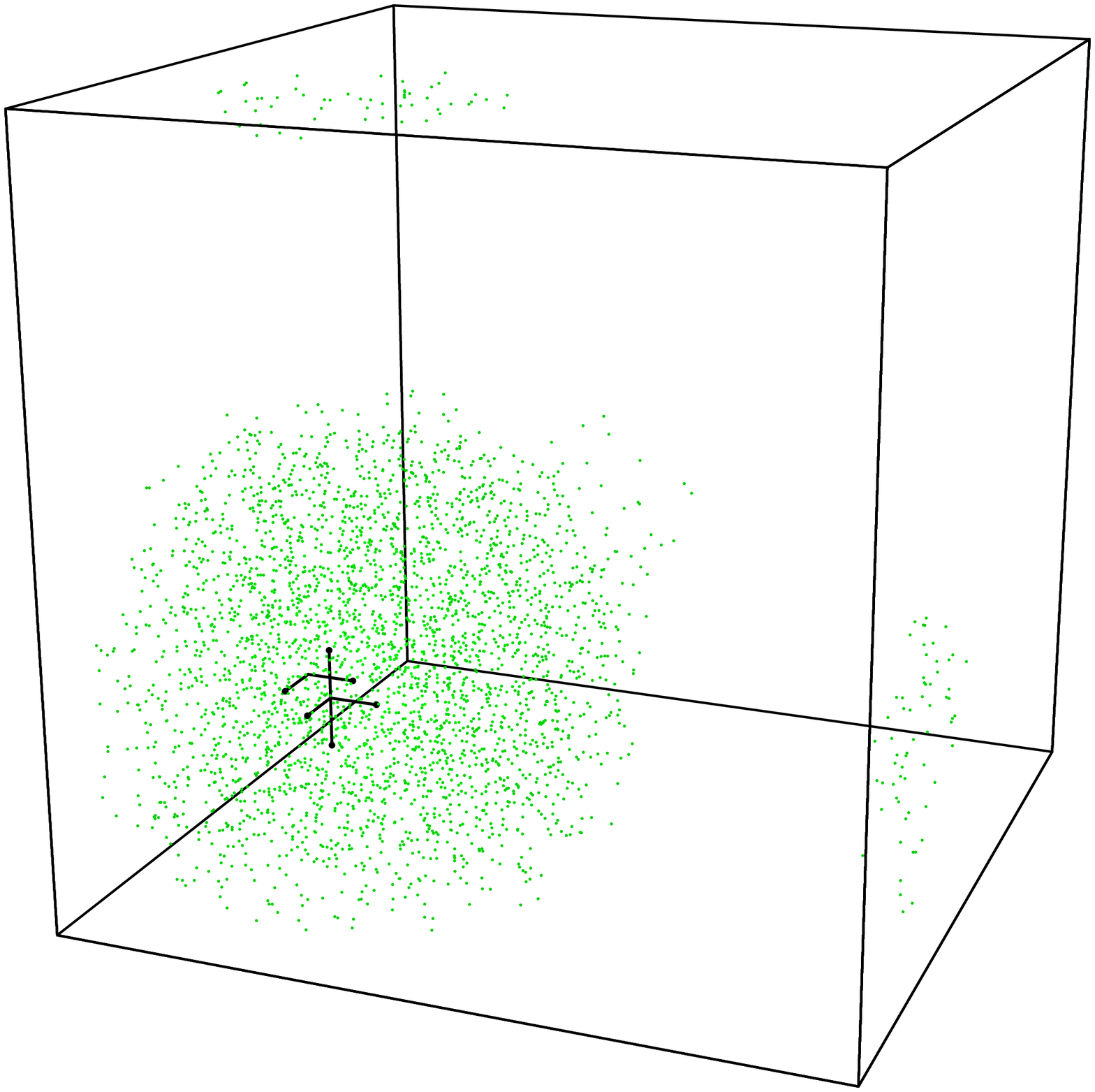} \\
(c) & (d)   \vspace{-0.3cm} \\ \\
\end{tabular}
\end{center}
\vspace{-0.5cm}
\caption{Same as Figure~2 but for the sphalerons with
$M_H=0.75~M_W$.}
\end{figure*}


\begin{thebibliography}{99}

\bibitem{KrLaScWi87} A.~S.~Kronfeld {\it et al.}, {\it Phys.~Lett.}
{\bf 198B} (1987) 516; A.~S.~Kronfeld, G.~Schierholz and U.~J.~Wiese,
{\it Nucl.~Phys.} {\bf B293} (1987) 461.

\bibitem{tH81}  G.~'t~Hooft, {\it Nucl.~Phys.} {\bf B190} [FS3]
(1981) 455.

\bibitem{Rev's} T.~Suzuki, {\it Nucl.~Phys.} {\bf B} {\it
(Proc.~Suppl.)} {\bf 30} (1993) 176; A.~Di~Giacomo, {\it Nucl.~Phys.}
{\bf B} {\it (Proc.~Suppl.)} {\bf 47} (1996) 136; M.~I.~Polikarpov,
{\it Nucl.~Phys.} {\bf B} {\it (Proc.~Suppl.)} {\bf 53} (1997) 134.

\bibitem{SuzukiBali} T.~Suzuki and I.~Yotsuyanagi, {\it Phys.~Rev},
{\bf D42} (1990) 4257; G.~Bali {\it et al.},
{\it Phys.~Rev.} {\bf D54} (1996) 2863.

\bibitem{Thermal}
K.~Kajantie {\it et al.},
{\it Nucl.~Phys.} {\bf B466} (1996) 189;
M.~G\"urtler {\it et al.},
{\it Nucl.~Phys.} {\bf B483} (1997) 383;
M.~G\"urtler, E.-M.~Ilgenfritz and A.~Schiller,
{\it Phys.~Rev.} {\bf D56} (1997) 3888.

\bibitem{Teper}
M.~Teper, {\it Phys.~Lett.} {\bf B311} (1993) 223.

\bibitem{Bornyakov}
V.~Bornyakov and R.~Grygoryev,
{\it Nucl.~Phys.} {\bf B} {\it (Proc.~Suppl.)}
{\bf 30} (1993) 576.

\bibitem{InProgress} M.~N.~Chernodub, F.~V.~Gubarev,
E.--M.~Ilgenfritz and A.~Schiller, in preparation.

\bibitem{DaHaNe74} R.~F.~Dashen, B.~Hasslacher and A.~Neveu, {\it
Phys.~Rev.} {\bf D10} (1974) 4138.

\bibitem{Ma83} N.~S.~Manton, {\it Phys.~Rev.} {\bf D28} (1983) 2019.

\bibitem{KlMa84} F.~R.~Klinkhamer and N.~S.~Manton, {\it Phys.~Rev.}
{\bf D30} (1984) 2212.

\bibitem{SphRev}
V.~A.~Rubakov and M.~E.~Shaposhnikov, {\it Usp.~Fiz.~Nauk}
{\bf 166}(1996) 493 ({\it Phys.~Usp.} {\bf 39} (1996) 461).

\bibitem{Na77} Y.~Nambu, {\it Nucl.~Phys.} {\bf B130} (1977) 505.

\bibitem{BaVaBu94} M.~Barriola, T.~Vachaspati and M.~Bucher,
{\it Phys.~Rev.} {\bf D50} (1994) 2819.

\bibitem{Hi94}
M.~Hindmarsh and M.~James, {\it Phys.~Rev.} {\bf D49} (1994) 6109;
M.~Hindmarsh, {\it Sintra Electroweak} (1994) 195, {\tt
hep-ph/9408241}.

\bibitem{PeBa95}
M.~Garcia~Perez and P.~van~Baal, {\it Nucl.~Phys.} {\bf B429} (1994) 451;
M.~Garcia~Perez and P.~van~Baal, {\it Nucl.~Phys.} {\bf B468} (1996) 277.

\bibitem{ANO}
A.~A.~Abrikosov, {\it Sov.~Phys.~JETP} {\bf 32} (1957) 1442;
H.~B.~Nielsen and P.~Olesen, {\it Nucl.~Phys.} {\bf B61} (1973) 45.

\bibitem{DGT}
T.~A.~DeGrand and D.~Toussaint, {\it Phys.~Rev.} {\bf D22} (1980) 2478.

\bibitem{ChPoZu94} M.~N.~Chernodub, M.~I.~Polikarpov and M.~A.~Zubkov,
{\it Nucl.~Phys.} {\bf B} {\it (Proc.Suppl.)} {\bf 34} (1994) 256.

\bibitem{tHPo74} G. `t~Hooft, {\it Nucl.~Phys.} {\bf B79} (1974) 276;
A.~M.~Polyakov, {\it JETP~Lett.} {\bf 20} (1974) 194.

\bibitem{bornyakov1} V.~G.~Bornyakov {\it et al.},
{\it Z.~f.~Phys.} {\bf C 42} (1989) 633.

\bibitem{VaBa69}
T.~Vachaspati and M.~Barriola, {\it Phys.~Rev.~Lett.}
{\bf 69} (1992) 1867.

\bibitem{private}
M.~Garcia~Perez, private communication.

\end{thebibliography}
\end{document}